# Stress, Strain, or Displacement? A Novel Machine Learning Based Framework to Predict Mixed Mode I/II Fracture Toughness

Amir Mohammad Mirzaei[1]

*Faculty of Engineering and Applied Science, Cranfield University, Cranfield MK43 0AL, UK*

**Abstract**

Accurate prediction of fracture toughness under complex loading conditions, like mixed mode I/II, is essential for reliable failure assessment. This paper aims to develop a machine learning framework for predicting fracture toughness and crack initiation angles by directly utilizing stress, strain, or displacement distributions represented by selected nodes as input features. Validation is conducted using experimental data across various mode mixities and specimen geometries for brittle materials. Among stress, strain, and displacement fields, it is shown that the stress-based features, when paired with Multilayer Perceptron models, achieve high predictive accuracy with $R^2$ scores exceeding 0.86 for fracture load predictions and 0.94 for angle predictions. A comparison with the Theory of Critical Distances (Generalized Maximum Tangential Stress) demonstrates the high accuracy of the framework. Furthermore, the impact of input parameter selections is studied, and it is demonstrated that advanced feature selection algorithms enable the framework to handle different ranges and densities of the representing field. The framework's performance was further validated for datasets with a limited number of data points and restricted mode mixities, where it maintained high accuracy. The proposed framework is computationally efficient and practical, and it operates without any supplementary post-processing steps, such as stress intensity factor calculations.

[1]Corresponding author. Email address: amir.mirzaei@crafield.ac.uk

# 1. Introduction

Fracture mechanics has historically been grounded in classical theories that provide analytical solutions for crack behavior under idealized conditions [1]. Among these, Griffith's energy criterion [2] established a foundational framework by balancing the elastic strain energy released during crack propagation with the energy required to create new surfaces. Later, the Maximum Tangential Stress (MTS) criterion [3] was introduced and became a critical tool for predicting fracture load and crack initiation angle under mixed mode loading conditions. Some years later, based on the energy distribution around the crack tip, the Strain Energy Density (SED) criterion [4] was proposed, which was also able to predict both fracture load and crack direction. Again, for mixed mode loading, a less fortunate model, called Maximum Strain Criterion [5] was proposed that employs the tangential strain component (similar to MTS) to predict mixed mode fracture. While classical fracture mechanics approaches provide clear, physics-based guidelines for predicting fracture, they typically rely on simplified assumptions that limit their range of applicability. This challenge grows when dealing with multiphysics problems, interacting failure mechanisms [6], or crack propagation in complex structures [7]. In recent years, machine learning (ML) has demonstrated high predictive accuracy in a variety of such cases, and we will briefly discuss some key contributions in this section. Although classical approaches often require parameter calibration, they still hinge on theoretical assumptions. By relying on observed patterns, machine learning models can 'learn' complex relationships that may not be captured by a single fracture criterion. This justifies the attempts to develop machine learning algorithms, beginning with simple problems and progressively advancing toward more complex fracture scenarios.

Recent advancements in ML have yielded remarkable progress in addressing mixed mode fracture problems, particularly through artificial neural networks (ANNs). For example, ANN-based models were successfully used to predict critical stress intensity factors and crack tip opening displacements in concrete [8]. These models outperformed traditional two-parameter models even in the presence of noisy datasets, achieving high accuracy and robustness. Another investigation [9] applied ANNs to predict fracture toughness and crack paths in materials influenced by micro-defects, with models trained on data derived from Distributed Dislocation Technique [10,11] and achieved high $R^2$ values for predictions. Furthermore, ANN-based approaches have been employed to address variables such as crack geometry, temperature, and biaxiality in fracture toughness prediction of aluminum alloys, demonstrating their ability to deliver rapid and cost-efficient fracture toughness predictions [12]. Beyond specific fracture problems, ANN models have been adapted to predict mixed mode fracture loads with high precision. For instance, an ANN-based criterion for mixed mode I/II fractures was developed that employs stress intensity factors and strain energy density as input features for fracture load predictions [13]. Another investigation employed stress intensity factors as well as geometric and material properties as inputs to predict mixed mode fracture load across diverse specimens and materials [14]. In addition, two-hidden-layer ANN architectures were applied to multi-crack scenarios, such as interacting cracks in aluminum alloys, which were effectively able to capture complex crack growth behaviors and directions [6]. These findings underscore the adaptability and robustness of ML approaches in addressing the non-linear interactions characteristic of mixed mode fracture mechanics. The applicability of ML models extends across material types and loading conditions. Regression trees and neural networks, for instance, have been applied to predict mode-I fracture toughness in brittle ceramics using experimental data from microcantilever tests. These models demonstrated superior generalization capabilities compared to empirical methods [15].

Support Vector Regression (SVR) has been applied to rocks under mode I and II loading conditions, achieving $R^2$ values of 0.73 and 0.77, respectively [16]. Another study [17] evaluated 12 ML models for predicting effective fracture toughness across multiple materials, and identified the Extreme Tree Regressor (ETR) as the most accurate. Practical tools such as graphical user interfaces (GUIs) have also been developed to enhance the usability of ML-based models by bridging the gap between theoretical advancements and real-world engineering applications. For instance, ANN and Extreme Gradient Boosting (XGBoost) models have been used to predict mode-I fracture toughness in concrete based on mix design parameters. These studies identified critical predictors such as cement dosage and notch height, achieving $R^2$ values as high as 0.90 [8,18]. Machine learning has also facilitated the calibration and validation of numerical models. For example, Gaussian Process Regression (GPR) has been integrated with finite element methods to automate failure model calibration in aerospace composites, eliminating the need for trial-and-error processes and significantly reducing computational costs [19].

Recent advancements in ML hybridization, particularly with physics-informed methods, have further enhanced their effectiveness in fracture analyses. Transfer learning–enhanced physics-informed neural networks (PINNs) have been employed for phase-field modeling of brittle fractures, achieving superior accuracy while significantly reducing computational costs [20]. Similarly, an extended physics-informed extreme learning machine (XPIELM) that incorporates singular factors derived from asymptotic expansions has been applied to address linear fracture problems which significantly improved displacement and stress intensity factor predictions [21]. Furthermore, ML-based approaches have been employed to address variational brittle fracture mechanics. For example, a model-free data-driven approach [22] replaced constitutive models with ML-based predictions and was able to accurately capture Griffith and R-curve behaviors. It is worth noting that despite the transformative potential, the widespread adoption of ML in fracture mechanics is not without challenges. Substantial datasets are often required for training, interpretability can be difficult in safety-critical domains, and overfitting remains a risk if hyperparameters and validation protocols are not carefully managed. Nevertheless, ensemble methods [23], regularization techniques [14], and hybrid approaches [20,21] have been successfully employed to address these challenges, ensuring the robustness and generalizability of ML models.

As reviewed, ML was employed to accurately capture complex data patterns to provide precise predictions for different fracture mechanics problems. The present study builds on the authors' previous work [24], where ML was coupled with stress, strain, or energy release rate data to predict fatigue life for notched components. The key idea was that different notch geometries result in distinct field distributions that ML can learn to predict the fatigue life. The present study extends the framework to predict mixed mode I/II fracture toughness and crack initiation angles for isotropic brittle and quasi-brittle materials. In simple terms, because field distribution of stress, strain, or displacement varies for different mode mixities and specimen types around the crack tip, the tangential components of these fields are used to train the ML models to find the most effective components to predict the fracture load and initiation angle. To clarify further, a surface plot of stress field around a crack for two different mode mixities are plotted in Fig. 1. Unlike most of the reviewed contributions, the proposed framework predicts both fracture load and initiation angle. By leveraging lightweight yet powerful tree-based ML algorithms and properly tuning ANN parameters, the framework addresses challenges such as limited datasets. If successful, the model can be considered efficient and straightforward, as it only requires a linear elastic analysis of the cracked sample without the need for additional postprocessing to calculate parameters such as stress intensity factors. In the next section, the machine learning algorithm designed for this study is explained in detail. Section 3 presents the experimental data used for validation, while Section 4 evaluates the accuracy and behavior of the framework for different mode mixities and specimen geometries. Additionally, that section presents different studies to demonstrate the effectiveness of the approach for different cases of data selection and availability.

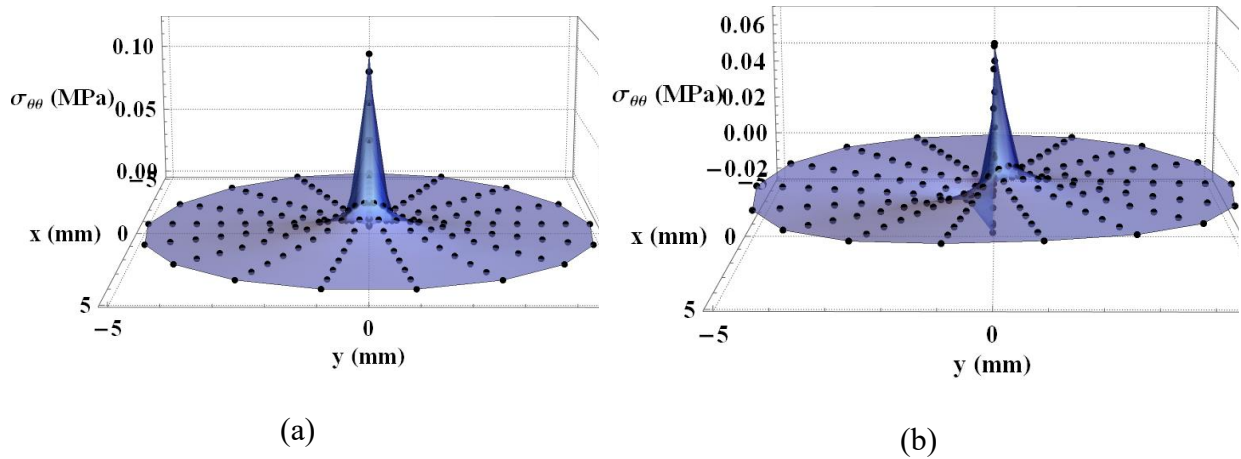

**Fig. 1.** Stress field distribution around a crack under (a) mode I loading and (b) mixed mode loading, with representative nodes.

## 2. Machine learning algorithm

This section presents the machine learning models used in this study, starting with the main steps, followed by an introduction to the employed neural network (NN) and tree-based models, and then explaining the input file in detail. Main equations related to each model are presented and explained to enhance understanding of each model and how the framework works.

Consider the dataset $D=\{X,Y\}$, which consists of a feature matrix $X \in \mathbb{R}^{N \times M}$, where $N$ is the number of samples and $M$ is the number of features. These features can be derived from stress, strain, or displacement field, sampled at nodal points around the crack tip. Instead of predicting multiple target variables at once, a single target variable is chosen for each training run. In other words, for fracture load prediction $Y=\{y_{force}\}$, and for fracture angle predictions $Y=\{y_{angle}\}$. Additionally, the dataset includes a categorical variable, Specimen Type, which, for example, can be a common name for different specimens with $n_{type}$ unique categories. This variable is used to distinguish between seen and unseen data.

The algorithm starts with a preprocessing step which includes outlier removal from the 'force' and 'angle' columns using an Interquartile Range (IQR) Rule and feature transformation using the Yeo–Johnson transformation [25]. This approach is a power-based technique that stabilizes variance and reduces skewness in both positive and negative-valued data. This method is used based on the physical problem, since the stress, strain, or displacement fields show a power-law dependency on radial distance, considering the asymptotic solution [26]. The Yeo–Johnson transformation is defined as [25]:

$$y(\lambda) = \begin{cases} \frac{(y+1)^{\lambda}-1}{\lambda}, & \text{if } y \geq 0, \ \lambda \neq 0, \\ \ln(y+1), & \text{if } y \geq 0, \ \lambda = 0, \\ -\frac{(-y+1)^{2-\lambda}-1}{2-\lambda}, & \text{if } y < 0, \ \lambda \neq 2, \\ -\ln(-y+1), & \text{if } y < 0, \ \lambda = 2. \end{cases} \quad (1)$$

where $\lambda$ is the transformation parameter, which controls the degree of nonlinearity.

The main algorithm begins with the dynamic identification of target variables, $T=\{y1, y2, ..., yT\}$, based on the dataset columns. Categorical variables such as Specimen Type are encoded using one-hot encoding [27], which maps the categorical values into binary vectors. If Specimen Type has $n_{type}$ unique categories, the encoding yields $n_{type}$ binary columns. The encoded categorical features are combined with the numerical features to create the augmented feature matrix $X_{augmented}=[X_{numerical}, C_{encoded}]$. In order to further reduce the effect of outliers and ensure numerical stability during model training, scaling transformations are used. Scaling transformations for both features and targets are applied separately to avoid data leakage, using robust scaling [28], which is expressed as:

$$S(X) = \frac{X - median(X)}{IQR(X)} \quad (2)$$

where $IQR(X)$ is the interquartile range of $X$. From a numerical perspective, robust scaling prevents extreme values in stress, strain, or displacement fields from disproportionately influencing the training process, thereby improving the convergence of optimization algorithms. From a physical standpoint, it normalizes features while preserving their relative magnitudes, which is essential for maintaining the integrity of stress gradients and crack-tip mechanics.

Unseen data grouping is employed to evaluate the generalization capability of the models. The Specimen Type categories are partitioned into $g$ disjoint groups such that:

$$T_{unseen} = \bigcup_{i=1}^{g} T_i \quad \text{and} \quad T_i \cap T_j = \emptyset \quad \forall i \neq j \tag{3}$$

For each group $T_i$, the associated samples are designated as the unseen dataset, $D_{unseen}$, while the remaining data make up the seen dataset, $D_{seen}$. In this paper, unseen data are selected across all the mode mixities and conditions related to a specific specimen type (geometry). In other words, all the information related to one geometry is considered unseen. This approach represents a more realistic scenario, as no training is performed using data from the targeted specimen geometry.

To effectively reduce the number of input features, feature selection is performed in three steps—Mutual Information (MI) analysis, Lasso-based feature reduction, and SHAP-based importance ranking—since stress, strain, or displacement fields around the crack tip may contain hundreds of nodes. MI [29] measures the dependency between each feature and target. Unlike correlation coefficients, MI accounts for complex interactions, such as those arising from localized crack-tip effects or stress gradients in mixed mode loading, making it particularly suitable for fracture mechanics problems. For a given feature $X_j$ and target $Y_t$, the MI is computed as:

$$MI(X_j, Y_t) = \iint p(x_j, y_t) \log \frac{p(x_j, y_t)}{p(x_j)p(y_t)} dx_j dy_t \tag{4}$$

where $p(x_j, y_t)$ is the joint probability distribution of $X_j$ and $Y_t$, and $p(x_j)$ and $p(y_t)$ are their marginal distributions. Features with the top 90% MI scores are retained as $F_{MI}$. Subsequently, Lasso regression [30] is employed to further refine the $F_{MI}$ set by selecting the most relevant features. Lasso applies an $L_1$-regularization penalty to shrink less important coefficients to zero, effectively performing feature selection while maintaining model interpretability. This is particularly advantageous for high-dimensional datasets, as it reduces redundancy and improves computational efficiency. The optimization objective for Lasso is:

$$\beta_t^* = \min_{\beta_t} \frac{1}{2N} \|Y_{:,t} - X_{MI}\beta_t\|_2^2 + \lambda \|\beta_t\|_1 \tag{5}$$

where $\beta_t$ represents the regression coefficients for the $t$-th target, and $\lambda$ controls the strength of regularization. Features with non-zero coefficients are retained as the $F_{Lasso}$. Finally, a Random Forest model is trained on $F_{Lasso}$ to compute SHAP (SHapley Additive exPlanations) values, which provide a measure of each feature's importance [31] (see section 4.3). SHAP is a game-theory-based approach [32] that assigns each feature a contribution value that reflects its impact on the model's prediction for a specific instance. It calculates these contributions by evaluating the effect of including or excluding a feature in all possible subsets of the feature set, ensuring that the importance is distributed fairly among features based on their marginal contributions to the prediction. For a given feature $X_j$ and $Y_t$, the SHAP value $\phi_{j,t}(x)$ is calculated as:

$$\phi_{j,t}(x) = \sum_{S \subseteq \{1,\ldots,m\}\setminus\{j\}} \frac{|S|!(M-|S|-1)!}{M!}[f_t(S \cup \{j\}) - f_t(S)] \quad (6)$$

where $f_t(S)$ is the model prediction using only the subset $S$ of features. Features with the highest mean absolute SHAP values are selected as $F_{SHAP}$ (final selected feature). This approach is specifically used to demonstrate the effectiveness of integrating machine learning with physical field distributions around the crack tip, and its results are presented in Section 4.3.

Subsequently, the models—MLP, RF, GB, and XGB—are trained iteratively using k-fold cross-validation [33] with $k=5$. In k-fold cross-validation, the dataset is divided into $k$ equal subsets, where $k-1$ subsets are used for training, and the remaining subset is used for validation; this process is repeated $k$ times, ensuring each subset serves as validation once. This method provides a robust estimation of model performance by reducing bias and variance, particularly in smaller datasets, while minimizing the risk of overfitting. This step also involves optimizing the machine learning hyperparameters to maximize predictive performance. To achieve this, the Tree-Structured Parzen Estimator (TPE) [34] is utilized. This approach systematically explores the hyperparameter space, evaluating different configurations to maximize the scoring function, such as $R^2$, ensuring the selected features contribute optimally to the model's predictions. Table 1 presents the hyperparameters and their corresponding ranges explored in this study. Once the models are trained, they are tested on unseen data, and the predictions are compared against true experimental values for each target variable. All the steps are detailed in Algorithm 1 to provide a structured overview of the feature selection, hyperparameter optimization, model training, and evaluation process.

**Algorithm 1: The regression framework used in this study.**

1. **Input:**

Dataset $D=\{X,Y\}$, where $X$ represents the feature matrix and $Y$ corresponds to the single target variable.

Categorical variable ***Specimen Type***.

Hyperparameters for feature selection, model training, and optimization:

Number of K-Folds $k$, number of unseen groups $g$, and search iterations $n_{itern}$.

Scalers $S_x$ and $S_y$ for features and targets, initialized as `RobustScaler`.

2. **Data Preprocessing:**

   Remove outliers from 'force' and 'angle' columns using the IQR-based method and apply the Yeo–Johnson transformation to features.

   Encode the categorical variable ***Specimen Type*** using one-hot encoding, adding $n_{type}$ binary columns to $X$.

3. **Unseen Data Grouping:**

   Partition ***Specimen Type*** categories into $g$ disjoint groups to simulate unseen data scenarios.

   For each group $g$:

   Assign $g_i$ as the unseen dataset $D_{unseen}$.

   Combine the remaining groups to form the seen dataset $D_{seen}$.

   Scale $D_{unseen}$ and $D_{seen}$ independently using $S_x$ and $S_y$.

4. **Feature Selection:**

   For each target variable $Y \in T$:

   **Step 1: Mutual Information Analysis (MI):**

   Select the top 90% features by MI to form $F_{MI}$.

   **Step 2: Lasso-Based Feature Reduction:**

   Retain non-zero features to form $F_{Lasso}$.

   **Step 3: SHAP Analysis:**

   Train a ***Random Forest Regressor*** model on $F_{Lasso}$ and compute SHAP values for interpretability.

   Rank features by mean absolute SHAP values and provide the top 5 features to form $F_{SHAP}$.

5. **Iterative Model Training and Evaluation:**

For each fold in K-Fold:

Split $D_{seen}$ into $D_{train}$ and $D_{val}$.

Train models (e.g., **MLP**, **RF…**) on $D_{train}$ using $F_{selected}$ as inputs.

Optimize hyperparameters using a Tree-Structured Parzen Estimator (TPE).

Validate models on $D_{val}$ and record evaluation metrics for each fold.

6. **Unseen Data Prediction and Aggregation:**

    Predict $T_{unseen}$ from $X_{unseen}$ using trained models.

7. **Evaluation and Visualization:**

    Compute metrics **{$R^2$, MAE, MAPE}** for unseen data predictions.

8. **Repeat Steps 4–8 for Each Group $g_i$.**

9. **Output:**

    Optimized models, feature importance rankings, unseen prediction metrics, and performance plots.

Having the structure of the framework, in the following the machine learning models used in this study are introduced.

**2.1. Multi-Layer Perceptron (MLP)**

The Multi-Layer Perceptron (MLP) is a fully connected feedforward neural network model that maps input features to target variables using multiple layers of weighted connections [35,36]. It employs non-linear activation functions to capture complex relationships in the data which makes it a powerful tool for predictive tasks involving high-dimensional feature spaces. The MLP implemented in this study has an input layer, two hidden layers, and an output layer.

The network accepts an input feature vector $\mathbf{x} \in \mathbb{R}^M$, where $M$ is the number of input features, and outputs predictions $\mathbf{y} \in \mathbb{R}^T$, where $T$ is the number of target variables. The forward pass through the network can be expressed as:

$$\begin{aligned}
&\mathbf{h}_1 = tanh(\mathbf{W}_1 \mathbf{x} + \mathbf{b}_1), \quad \mathbf{h}_1^{dropout} = \text{Dropout}(\mathbf{h}_1, p), \\
&\mathbf{h}_2 = tanh(\mathbf{W}_2 \mathbf{h}_1^{dropout} + \mathbf{b}_2), \quad \mathbf{h}_2^{dropout} = \text{Dropout}(\mathbf{h}_2, p), \\
&\mathbf{y} = \mathbf{W}_3 \mathbf{h}_2^{dropout} + \mathbf{b}_3.
\end{aligned} \quad (7)$$

Here, $\mathbf{W}_i$ and $\mathbf{b}_i$ are the weight matrices and biases for layer $i$, and tanh($x$) is the hyperbolic tangent activation function. This activation function maps inputs to [−1,1], which introduces non-linearity and enables the network to model complex patterns. Dropout regularization [37] is applied after each hidden layer to prevent overfitting by randomly deactivating a fraction $p$ of neurons during training:

$$\mathbf{h}_1^{dropout} = \mathbf{h}_1 \odot \mathbf{m}, \quad \mathbf{m} \sim Bernoulli(1-p) \quad (8)$$

where $\odot$ represents element-wise multiplication and $\mathbf{m}$ is a random binary mask. The output layer, which applies no activation, allows the network to produce continuous predictions suitable for regression.

Training minimizes the Mean Squared Error (MSE) loss function, which is defined as:

$$\text{MSE} = \frac{1}{N} \sum_{i=1}^{N} \| \mathbf{y}_{i,t} - \mathbf{y}_{i,p} \|^2 \quad (9)$$

where $\mathbf{y}_{i,t}$ and $\mathbf{y}_{i,p}$ are the true and predicted targets, and $N$ is the number of samples. The Adam optimizer updates weights efficiently, combining momentum and adaptive learning rates:

$$\theta_t = \theta_{t-1} - \alpha \frac{\mathbf{m}_t}{\sqrt{\mathbf{v}_t} + \varepsilon} \quad (10)$$

where $\mathbf{m}_t$ and $\mathbf{v}_t$ are bias-corrected estimates of gradient moments, $\alpha$ is the learning rate, and $\varepsilon$ ensures numerical stability. Hyperparameters such as the number of neurons in hidden layers ($h_1, h_2$), dropout rate ($p$), and learning rate ($\alpha$) are tuned using the Tree-Structured Parzen Estimator; see Table 1.

## 2.2. Random Forest Regressor

Random Forest (RF) is an ensemble learning method designed to enhance prediction accuracy and robustness by combining the outputs of multiple decision trees [38,39]. It is based on the concept of bagging (Bootstrap Aggregating [40]), where each tree is trained on a random subset of the data, in terms of both samples and features. The final prediction is obtained by averaging (for regression) or majority voting (for classification) across all the individual tree predictions. RF offers two key advantages: it significantly reduces the variance of predictions compared to a single decision tree, improving generalization to unseen data and effectively capturing complex, non-linear relationships within the dataset. This makes it particularly suitable for this problem, where interactions among different features in the stress, strain, or displacement field often exhibit non-linear behavior.

Each tree in the forest is constructed independently, and the final prediction for a given input is the aggregation of predictions from all trees. Mathematically, the prediction can be expressed as:

$$y_{i,p} = \frac{1}{K} \sum_{k=1}^{K} f_k(x_i), \quad f_k \in F \tag{11}$$

where $f_k$ is the prediction from the $k$-th tree, and $F$ is the space of regression trees. The term $K$ represents the total number of trees in the forest.

The training process implicitly minimizes prediction error by reducing variance across the ensemble:

$$J(\Theta) = \sum_{i=1}^{K} \ell(y_{i,t}, y_{i,p}) \tag{12}$$

where Θ represents the parameters of the forest. While RF lacks explicit regularization terms, the bagging process and averaging inherently reduce overfitting. Additionally, RF handles high-dimensional data effectively and can model complex, non-linear relationships.

The key hyperparameters considered for the RF model in this study include the number of estimators ($K$), which defines the total number of decision trees in the ensemble. In general, larger values improve stability and accuracy at the cost of higher computational effort. The maximum tree depth limits how deep each tree can grow, controlling model complexity to balance the risks of underfitting and overfitting. The minimum number of samples required to split a node regulates tree complexity by preventing splits on small sample sizes to enhance generalization.

### 2.3. Gradient Boosting Regressor

The Gradient Boosting (GB) model is a tree-based ensemble method [41,42] that captures non-linear relationships by sequentially combining weak learners [43], where each learner corrects the residuals of the previous one. The prediction can be computed as:

$$y_{i,p} = \sum_{k=1}^{K} f_k(x_i), \quad f_k \in F \tag{13}$$

Here, $F$ represents the space of regression trees, and $f_k$ is the $k$-th tree in the sequence.

Each tree is trained to minimize the negative gradient of the loss function at iteration $t$:

$$g_t = -\frac{\partial \ell(y_t, y_p)}{\partial y_p} \tag{14}$$

The predictions are updated as:

$$y_{i,p}^t = y_{i,p}^{t-1} + \eta f_t(x_i) \tag{15}$$

where $\eta$ is the learning rate that scales the contribution of each tree to the final prediction.

For GB, the key hyperparameters include the number of estimators (K), the maximum tree depth, and the learning rate ($\eta$).

**2.4. XGBoost Regressor**

Building upon the framework of GB, eXtreme Gradient Boosting (XGBoost) enhances the methodology by incorporating advanced optimization techniques and explicit regularization [31,44]. This results in a model that is not only highly accurate but also computationally efficient and scalable. While Gradient Boosting in Scikit-learn [45] relies on structural constraints for regularization, XGBoost integrates these constraints directly into the objective function, offering finer control over model complexity.

XGBoost constructs an additive model similar to Gradient Boosting, however, the objective function in XGBoost combines the loss term and a regularization term:

$$J(\Theta) = \sum_{i=1}^{K} \ell(y_{i,t}, y_{i,p}) + \sum_{i=1}^{K} \Omega(f_k) \tag{16}$$

$J(\Theta)$ represents the objective function optimized during the training process. This function combines the loss term ($\ell$) and the regularization term ($\Omega(f_k) = (1/2) \lambda \|w_k\|^2 + \gamma T_k$), to penalize the complexity of the tree $f_k$, with $w_k$ representing the leaf weights, $\lambda$ controls $L2$ regularization, $\gamma$ penalizes the addition of new leaf nodes, and $T_k$ is the number of leaf nodes. This regularization term helps to prevent overfitting and ensures that the model generalizes well to unseen data.

Unlike Scikit-learn's Gradient Boosting, XGBoost optimizes the loss function using a second-order Taylor expansion. At iteration $t$, the objective is approximated as:

$$J(\Theta) \approx \sum_{i=1}^{N} \left[ g_i f_k(x_i) + \frac{1}{2} h_i f_k^2(x_i) \right] + \Omega(f_k) \tag{17}$$

where

$$g_i^t = \frac{\partial \ell\left(y_{i,t}^t, y_{i,p}^{t-1}\right)}{\partial y_{i,p}^{t-1}} \tag{18}$$

$$h_i^t = \frac{\partial^2 \ell\left(y_{i,t}^t, y_{i,p}^{t-1}\right)}{\partial \left(y_{i,p}^{t-1}\right)^2} \tag{19}$$

Each tree in the ensemble focuses on correcting the residuals from previous iterations, improving the overall prediction accuracy.

The hyperparameters and their ranges for XGB, are presented in Table 1.

**Table 1. Hyperparameters and their corresponding ranges for each machine learning model used in this study.**

| Model | Hyperparameters |
|---|---|
| Multi-Layer Perceptron | learning_rate: (0.001–0.05)<br>dropout_rate: (0.0–0.2)<br>hidden_units_1: (16–128)<br>hidden_units_2: (16–128) |
| Random Forest | n_estimators: (50–300)<br>max_depth: (3–15)<br>min_samples_split: (2–10)<br>min_samples_leaf: (1–5) |
| Gradient Boosting | n_estimators: (50–300)<br>max_depth: (3–15)<br>learning_rate (0.001–0.1)<br>subsample: (0.7-1.0)<br>min_samples_split: (2-10) |
| XGBoost | n_estimators: (50–300)<br>max_depth: (3–15)<br>learning_rate (0.001–0.1)<br>subsample: (0.7-1.0)<br>colsample_bytree: (0.7-1.0) |

As can be seen the selected models cover a wide range of learning paradigms—neural networks (MLP), bagging-based ensemble learning (RF), boosting-based sequential learning (GB), and optimized boosting (XGBoost). This ensures that the framework is robust to different feature dependencies, data distributions, and levels of nonlinearity. Furthermore, comparing these models enables a systematic evaluation of their effectiveness in predicting fracture loads and crack initiation angles, which can support future methodological advancements in data-driven fracture mechanics.

## 2.5. Input File

To calculate the stress, strain, and displacement fields, the commercial finite element (FE) software package Abaqus® was employed. The specimens were modelled under 2D plane-stress conditions using a linear elastic material model, with a highly refined mesh around the crack tip to accurately capture the stress singularity. The discretization was performed using 8-node biquadratic elements, and a convergence analysis was conducted to ensure the numerical results were independent of the mesh size. The simulations were performed under a remote force with an amplitude of 1 N. A representation of the meshed samples, as well as the boundary conditions, is presented in Fig. 2. Further details are not provided, as they fall outside the scope of this research; for more information, please refer to [24].

Based on the numerical analyses, the tangential (hoop) components of the stress, strain, and displacement fields were extracted up to a sufficient distance (discussed in Section 4.4), where the stress field, for instance, reached a nearly constant value. Next, Williams' analytical solution for the fields was fitted to the data by including higher-order terms (up to four) to achieve an accurate representation of the field [26,46]. Next, the field was extracted at specific fixed points across all cases. Since the number of nodes representing the field is flexible and a key parameter (discussed in Section 4.4), this process was designed to minimize mesh size effects and ensure accurate and efficient input selection. The process began by creating a structured table to evaluate a function over a defined range of radial and angular coordinates, both of which were evenly spaced between the specified start and end points (details are discussed in Section 4.4). For each radial node, the function was evaluated at every angular node, producing a nested structure where each row corresponded to a fixed radial position and contained computed values for all angular positions. The table was then flattened into a one-dimensional array, maintaining the order of evaluations. In this array, the function values were listed sequentially, first progressing through all angular evaluations for the smallest radial position, followed by the next radial position, and continuing in this manner until all radial and angular combinations were included.

Therefore, the input file consisted of the fracture load and initiation angle as targets, the abbreviation of the specimen geometry to help with categorization, and depending on the analysis, one of the corresponding fields—stress, strain, or displacement—was included under a load amplitude of 1 N. Each field is unique to a specific mode mixity and specimen, effectively distinguishing between various cracked configurations. Overall, the proposed framework is easy to implement, as it requires only linear elastic analyses for each configuration and does not involve any postprocessing to determine fracture parameters such as $K_I$, $T$-stress [47], or others.

## 3. Experimental data

In order to validate the proposed framework, two sets of experimental data are combined. The first set interestingly contains 4 different cracked specimens under 6 different mode mixities (from pure mode I to pure mode II) [48], and the second set contains 8 different mode mixity conditions [49]. All specimens were made from PMMA, a material characterized by its brittle, homogeneous, and isotropic behavior. It is worth emphasizing that all experiments were conducted within the same laboratory. The first dataset included test specimens in both triangular bend and semicircular bend configurations. These consisted of asymmetric and symmetric edge-cracked triangular specimens (AECT and SECT) as well as asymmetric and symmetric edge-cracked semicircular specimens (AECS and SECS). The second dataset, however, only included SECS configurations. The mode mixity in these specimens was controlled by varying the crack inclination angle ($α$) in SECT and SECS specimens or by adjusting the loading support distance (S2) in AECT and AECS specimens. All specimens were cut from a 5 mm thick PMMA sheet. For the first set, each test configuration was repeated three times, whereas in the second set, some cases were repeated more than three times. In total, the experiments produced 100 test data points. In Table 2, the geometry and loading details of the fracture toughness tests employed in this study are presented. For more details, see [48,49].

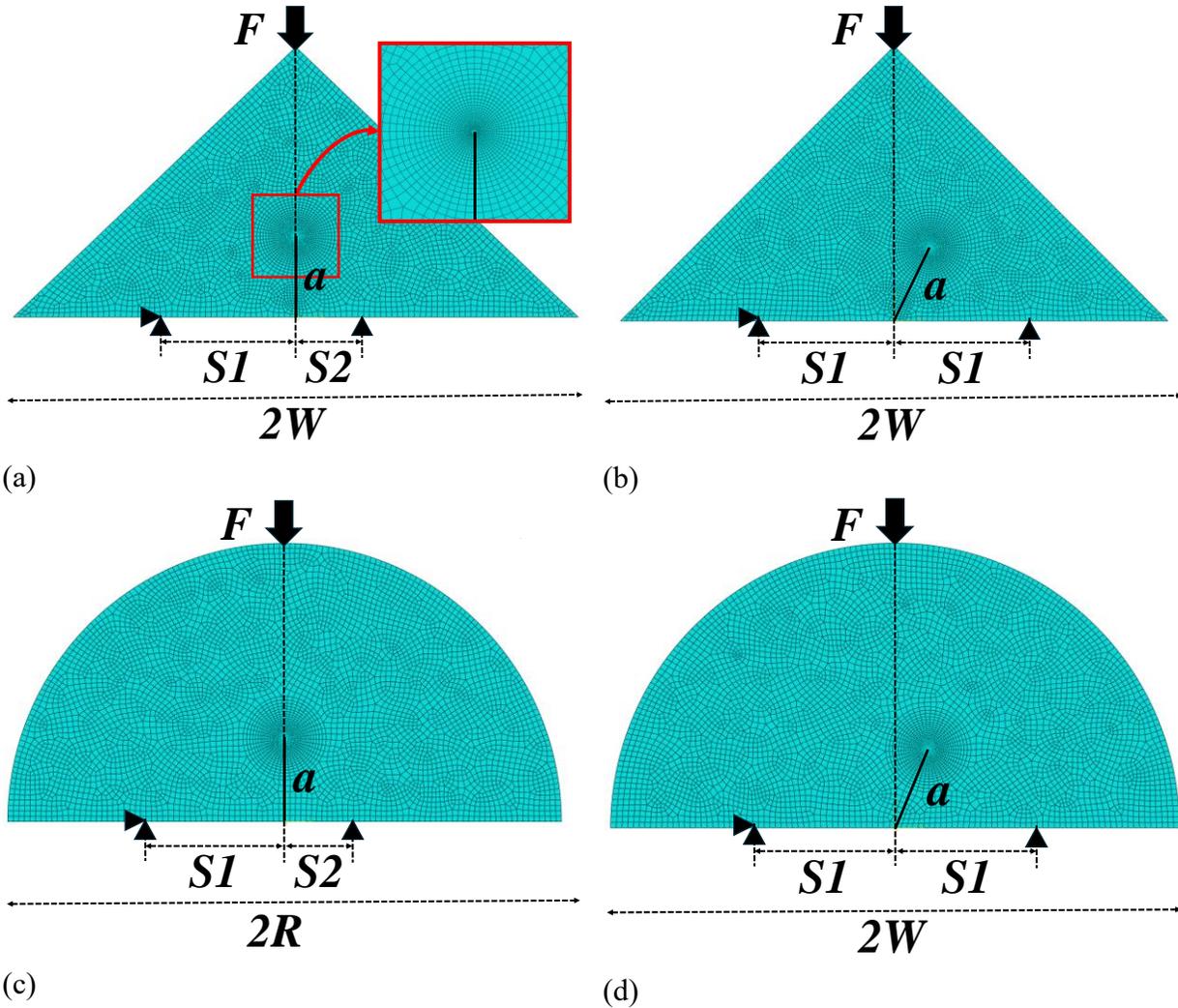

Fig. 2. Specimens used for mixed mode I/II validation of the proposed framework: (a) AECT, (b) SECT, (c) AECS, and (d) SECS.

Table 2. Geometrical and loading specifications of the employed fracture toughness tests.

| Specimen | R (mm) | W (mm) | a (mm) | t (mm) | S1 (mm) |
|---|---|---|---|---|---|
| AECT | - | 50 | 15 | 5 | 20 |
| SECT | - | 50 | 15 | 5 | 20 |
| AECS | 50 | - | 15 | 5 | 20 |
| SECS1 | 50 | - | 15 | 5 | 20 |

| SECS2 | 50 | - | 15 | 5 | 21.5 |

Since the Generalized Maximum Tangential Stress (GMTS) criterion [47], also known as the Theory of Critical Distances (TCD) [50,51], was used to evaluate the experimental data in the original research papers, it is also employed here as a benchmark for a comprehensive comparison of the framework's performance, see Section 4.2.

The GMTS criterion postulates:

1. Brittle fracture initiation occurs along the direction $\theta_0$, where the tangential stress evaluated at the critical distance ($r_c$) from the crack tip is maximized, Eq. 20.

2. Mixed mode fracture occurs when the tangential stress along $\theta_0$, at the critical distance ($r_c$) reaches the critical stress ($\sigma_{\theta\theta c}$), Eq. 21.

Mathematically, this can be expressed as:

$$\left.\frac{\partial \sigma_{\theta\theta}}{\partial \theta}\right|_{\theta=\theta_0} = 0 \tag{20}$$

$$\sigma_{\theta\theta}(r=r_c, \theta=\theta_0) = \sigma_{\theta\theta,c} = \sigma_t \tag{21}$$

Both $r_c$ and $\sigma_{\theta\theta c}$ are material properties. For brittle materials such as PMMA, $\sigma_{\theta\theta c}$ can be determined using the tensile strength ($\sigma_t$). Taylor et al. [52] proposed the following relationship for estimating $r_c$ in PMMA:

$$r_c = \frac{1}{2\pi}\left(\frac{K_{Ic}}{\sigma_t}\right)^2 \tag{22}$$

where $K_{Ic}$ represents the mode I fracture toughness.

The GMTS criterion incorporates the first non-singular stress component (*T*-stress) from Williams' solution for stress field calculation around cracks [53], which improves the accuracy of stress calculations near the crack tip compared to the traditional Maximum Tangential Stress (MTS) criterion [3].

Two distinct values for $r_c$ were reported in the experimental studies used in this investigation. In the first study, the strength and fracture toughness were measured and mentioned explicitly as $\sigma_t = $ 75 MPa and $K_{Ic} = $ 1.88 MPa√m resulting in $r_c = $ 0.1 mm. In the second one, only the mode I fracture toughness of 2.13 MPa√m for PMMA was reported, and no explicit mechanical properties such as tensile strength were provided. In this paper, instead of deriving $r_c$ using material properties, it was determined through curve fitting to best match the experimental data, yielding a value of $r_c = 0.065$ mm.

## 4. Results and discussion

### 4.1. Predictions across different components

In this section, the fracture load and initiation angle predictions of the proposed framework for the experimental data, using different components (stress, strain, or displacement) and ML models (MLP, RF, GB, XGB), are presented. To prepare the input file for the proposed framework, stress, strain, or displacement components were selected over a radial range starting at $r_c/5=0.02$ and extending to 5 mm. Along this radial distance, 10 points were sampled, while 18 points were selected along the hoop axis, resulting in a total of 180 data points. For each specimen, its corresponding abbreviation (SECT, AECT, AECS, SECS1, and SECS2) was also recorded to facilitate data grouping. As mentioned before, in this study, unseen data are selected from all mode mixities and conditions for a specific specimen geometry. This provides a realistic assessment, as no training data from the selected geometry is used. In order to compare the different cases and clearly demonstrate the accuracy of the framework, the analysis utilizes three key performance metrics: the coefficient of determination ($R^2$), mean absolute error (MAE), and mean absolute percentage error (MAPE). Before proceeding, it is useful to recall the key characteristics of each error metric. In general, MAE is given the highest priority because it directly quantifies the absolute error in units of the target, which is crucial for assessing structural safety and reliability. $R^2$ follows, as it reflects how well the model captures the variance in the data, ensuring a strong fit between predictions and observations. While MAPE is important for standardizing errors across datasets, it holds less weight in this context since absolute force errors are more practically relevant than percentage-based errors. It is important to note that MAPE can become misleadingly large when the values are very small, especially for fracture angles in mode I in this study. For the sake of completeness, the formula for each metric is presented below:

$$R^2 = 1 - \frac{\sum_{1}^{n}(y_{i,t} - y_{i,p})^2}{\sum_{1}^{n}(y_{i,t} - y_{i,mean})^2} \tag{23}$$

$$MSE = \frac{1}{n}\sum_{1}^{n}|y_{i,t} - y_{i,p}| \tag{24}$$

$$MAPE = \frac{100}{n}\sum_{1}^{n}\left|\frac{y_{i,t} - y_{i,p}}{y_{i,t}}\right| \tag{25}$$

The predictions for both fracture load and initiation angle (denoted as Force and Angle) are presented in Table 3, when the stress field is used as the feature. Similarly,

Table 4 presents the results when the strain field is used, and Table 5 provides the corresponding results for the displacement field. It is important to emphasize that these results are based on real experimental data, which naturally includes scatter and variability. Real experiments are harder to predict due to measurement noise, variations in experimental conditions, and other uncertainties. Moreover, the performance metrics are calculated after applying the inverse transformations, rather than directly in the transformed domain.

Table 3. Predictions for fracture load (Force) and initiation angle (Angle) using the stress field as the feature.

| Target | Model | $R^2$ | MAE | MAPE |
|---|---|---|---|---|
| Force | MLP | 0.86 | 0.19 | 6.7 |
|  | RF | 0.81 | 0.23 | 7.8 |
|  | GB | 0.80 | 0.24 | 8.2 |
|  | XGB | 0.83 | 0.22 | 8.1 |

|  | MLP | 0.94 | 5.19 | 67.0 |
| --- | --- | --- | --- | --- |
| Angle | RF | 0.94 | 5.16 | 44.5 |
|  | GB | 0.89 | 6.87 | 20.4 |
|  | XGB | 0.83 | 9.17 | 230.5 |

**Table 4. Predictions for fracture load (Force) and initiation angle (Angle) using the strain field as the feature.**

| Target | Model | $R^2$ | MAE | MAPE |
| --- | --- | --- | --- | --- |
| Force | MLP | 0.77 | 0.25 | 8.7 |
|  | RF | 0.76 | 0.26 | 8.9 |
|  | GB | 0.76 | 0.25 | 8.5 |
|  | XGB | 0.74 | 0.27 | 9.5 |
| Angle | MLP | 0.94 | 5.17 | 69.7 |
|  | RF | 0.92 | 5.73 | 34.6 |
|  | GB | 0.93 | 5.17 | 19.4 |
|  | XGB | 0.90 | 8.02 | 233.9 |

**Table 5. Predictions for fracture load (Force) and initiation angle (Angle) using the displacement field as the feature.**

| Target | Model | $R^2$ | MAE | MAPE |
| --- | --- | --- | --- | --- |
| Force | MLP | 0.87 | 0.19 | 6.5 |
|  | RF | 0.85 | 0.20 | 7.1 |
|  | GB | 0.86 | 0.20 | 6.9 |
|  | XGB | 0.82 | 0.23 | 8.1 |
| Angle | MLP | 0.76 | 10.07 | 49.3 |
|  | RF | 0.72 | 10.89 | 30.5 |
|  | GB | 0.68 | 12.03 | 27.1 |
|  | XGB | 0.64 | 12.60 | 223.4 |

The results suggested several consistent trends across the feature sets and targets. Stress-derived features appeared to provide more reliable predictions for both fracture load and initiation angle across all models, especially when used with the MLP model. Considering stress-based features for fracture load prediction, the MLP model achieved the lowest MAE (0.19 kN) and the highest R² (0.86) among all models. For fracture angle, the MLP model demonstrated strong performance with an R² of 0.94 and an MAE of 5.19° which was similar to that of RF in this case. These findings highlight the potential of stress features in capturing the mechanical behavior associated with fracture processes. Stress components directly represent the force distribution that drives crack initiation and growth. This leads to a powerful computational tool for fracture prediction, which also aligns well with the underlying physical behavior.

Strain-based features showed promising performance for predicting fracture angle but demonstrated a noticeable decline in accuracy for fracture load prediction. For load prediction, the MLP model achieved an $R^2$ of 0.77 and an MAE of 0.25 kN, outperforming all other models. Additionally, for angle prediction, MLP achieved an $R^2$ of 0.93 and MAE of 5.17°, closely matching the performance of the GB model.

Compared to strain-based features, displacement-derived features consistently demonstrated high accuracy for fracture load prediction but lower predictive ability for fracture angle prediction. For load prediction, the MLP model performed slightly better than the other models across all the features, achieving an $R^2$ of 0.87 and an MAE of 0.19 kN. This indicates that while displacement features are less directly tied to physical mechanisms compared to stress or strain, they can still effectively capture the global behavior of the system. However, for angle prediction, displacement features were significantly less effective. The $R^2$ of MLP dropped to 0.76, and its MAE increased to 10.07°. This decline in accuracy underscores the limitations of displacement features in capturing localized mechanical responses. The displacement field primarily represents the cumulative effects of stress and strain but lacks the fine resolution needed to describe localized phenomena, such as crack tip behavior, which is crucial for accurately predicting fracture angles. On the other hand, lower accuracy in fracture angle predictions might be due to the hoop component of displacement not being a good candidate for fracture angle prediction, although it performs reasonably well for fracture load estimations. It is worth mentioning that the tangential displacement component is calculated after eliminating the rigid body motions of the sample, and including these components in the calculation would result in slightly lower accuracy.

The superior performance of the MLP model across most cases suggests that it is particularly well-suited for this problem. A key reason why the MLP model outperforms other models in this problem lies in its ability to capture complex, nonlinear relationships between the input features and fracture responses. Unlike tree-based models (RF, GB, XGB), which rely on hierarchical, axis-aligned splits and may struggle to model smooth functions in high-dimensional spaces due to their piecewise-constant nature, MLP leverages its multi-layer architecture with nonlinear activation functions to learn intricate mappings between stress, strain, or displacement fields and the corresponding fracture load and angle. Additionally, MLP's backpropagation-based optimization allows it to adjust weights continuously, rather than relying on discrete rule-based splits which enables it to generalize better across different stress and strain distributions. Furthermore, the MLP's ability to capture both global trends and local variations, considering the network's architecture, the input representation, and the training data, makes it particularly well-suited for predicting fracture initiation angles, where even minor deviations in stress or strain concentrations can lead to substantial differences in crack direction. In contrast, gradient boosting models (GB, XGB) rely on additive corrections from weak learners, which can make them sensitive to noise and struggle with extrapolating beyond the training distribution. Finally, RF, as an ensemble of decision trees, provided the second-best results across all input feature sets. This can be attributed to its ability to aggregate multiple decision trees and effectively reduce overfitting while capturing key feature interactions. However, due to its independent tree construction and reliance on majority voting or averaging, RF lacks the fine-grained adaptability of MLP.

Beyond predictive accuracy, these findings highlight the importance of stress-based inputs for fracture analysis. The strong performance of stress features suggests that future studies focusing on similar problems may benefit from prioritizing stress-driven inputs. It is worth noting that a previous study by the author [24] showed similar results, where stress-based features, in general, performed best for fatigue life estimation across three different materials. However, it is important to note that these analyses are specific to materials that exhibit stress-based failure, and conclusions drawn here may not apply to strain-controlled failure materials such as bone [54].

Although predictions based on the displacement or strain field were generally less accurate than those based on the stress field, they still achieved a reasonably high level of accuracy. This highlights a noteworthy application of the proposed framework in its compatibility with displacement or strain fields, which can be directly measured using techniques such as digital image correlation (DIC). By integrating with DIC systems, the framework offers a practical method for real-time, contact-free fracture assessments without post-processing or prior knowledge of the entire geometry or thickness (if failure mode does not change). It uses localized strain or displacement data to predict critical fracture parameters, making it a reliable tool for monitoring complex structures and materials in real-world applications. This approach is especially useful in situations where traditional methods may not be feasible, providing a straightforward and efficient alternative for fracture analysis.

To better understand the accuracy of predictions across different specimens and loading conditions, the predictions by the MLP model for stress, strain, and displacement are visualized in Fig. 3 to Fig. 5. In these figures, part (a) illustrates the fracture load (Force) predictions, while part (b) presents the initiation angle predictions for all specimens. The solid black line represents the ideal prediction, serving as a benchmark for comparison. Furthermore, ±20% scatter bands are depicted using red dashed lines.

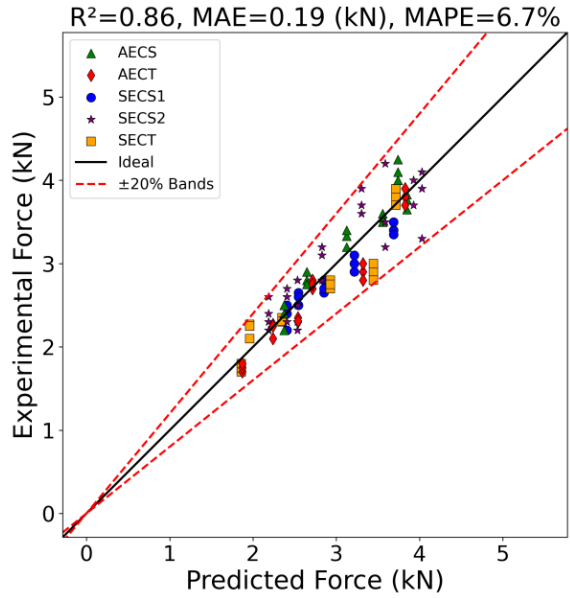
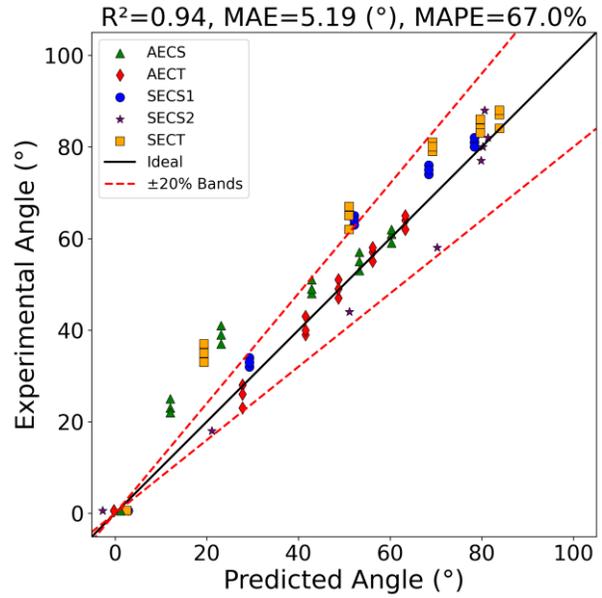

(a)                                         (b)

**Fig. 3.** Predictions of fracture load (a) and initiation angle (b) for different specimens using the stress field as input features. The solid black line represents the ideal prediction, while the red dashed lines indicate ±20% scatter bands.

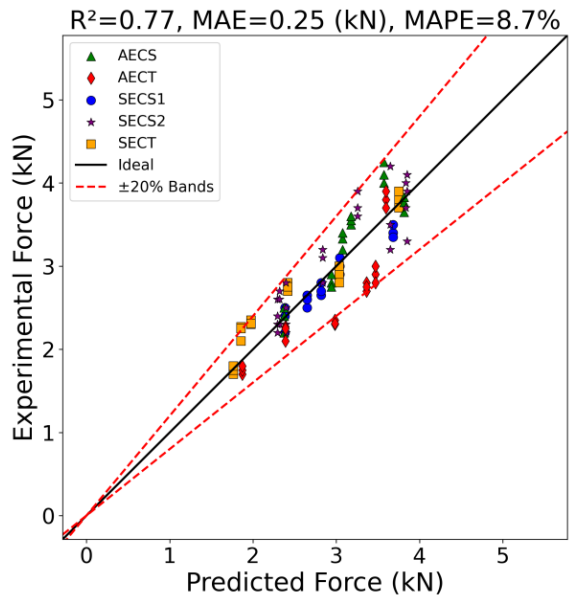
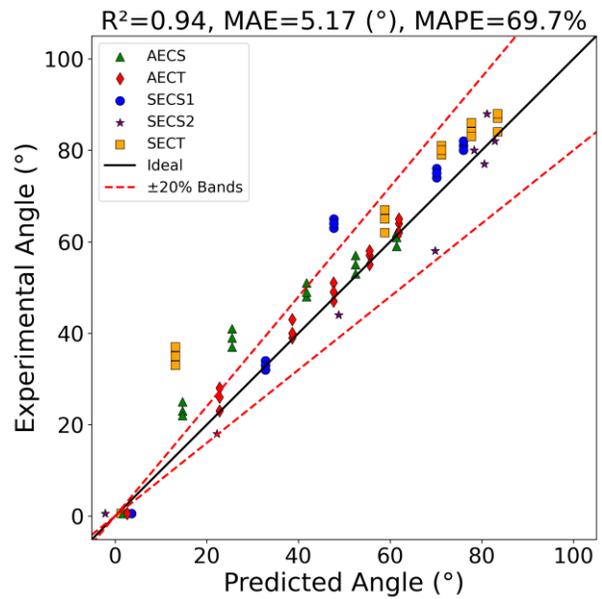

(a)                                         (b)

**Fig. 4. Predictions of fracture load (a) and initiation angle (b) for different specimens using the strain field as input features. The solid black line represents the ideal prediction, while the red dashed lines indicate ±20% scatter bands.**

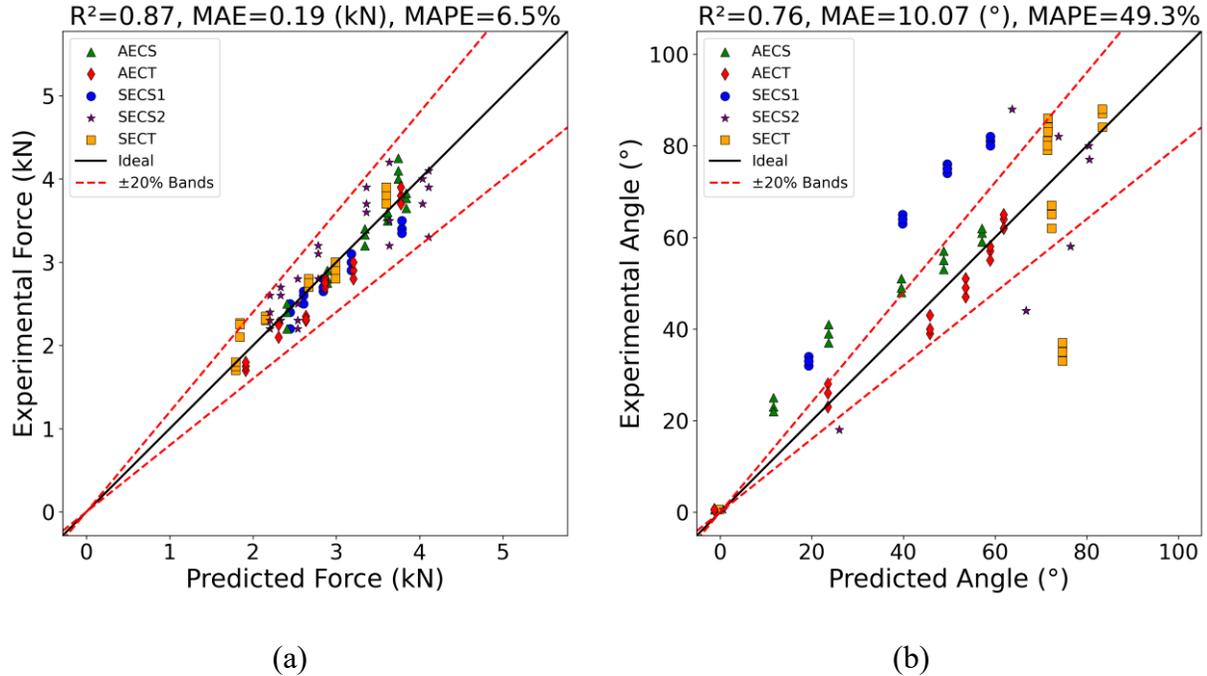

(a)          (b)

**Fig. 5. Predictions of fracture load (a) and initiation angle (b) for different specimens using the displacement field as input features. The solid black line represents the ideal prediction, while the red dashed lines indicate ±20% scatter bands.**

Considering the stress field as input (Fig. 3), the MLP model demonstrated excellent alignment with experimental data for force predictions, with all predictions falling within the ±20% scatter bands. The clustering of predictions around the ideal line suggests that stress-based features effectively capture the mechanics driving fracture loads. Angle predictions also showed strong agreement with experimental data, achieving a higher $R^2$ value than load predictions. However, some of the predictions for AECS and SECT are outside the scatter bands. On the other hand, the high MAPE value in this case is due to its sensitivity to small values. Even minor errors in predicting fracture angles under mode I conditions, where angles are close to zero, can lead to disproportionately large MAPE values.

Strain-based inputs, as shown in Fig. 4, performed comparably well, with force predictions aligning closely with experimental results and achieving lower errors across key metrics. Angle predictions also exhibited strong accuracy, with most points within the scatter bands, though the slightly higher MAPE indicates reduced relative accuracy. As shown in Fig. 5, displacement-based inputs demonstrated the best performance across all the inputs for force predictions with $R^2=0.87$; however, fracture angle predictions displayed more variability.

## 4.2. A comparison with GMTS

To better understand the predictive capability of the proposed framework and the complexity of the problem, predictions across different specimens and mode mixities are also calculated using GMTS, one of the most developed analytical models for fracture load and initiation angle. This comparison provides additional insight into how well the framework performs relative to established methods. For fracture load, Fig. 6 presents the results of GMTS in part (a), while part (b) presents the predictions from the MLP model using stress-based inputs. Similarly, for the initiation angle, Fig. 7 compares the predictions from GMTS to those of the proposed framework.

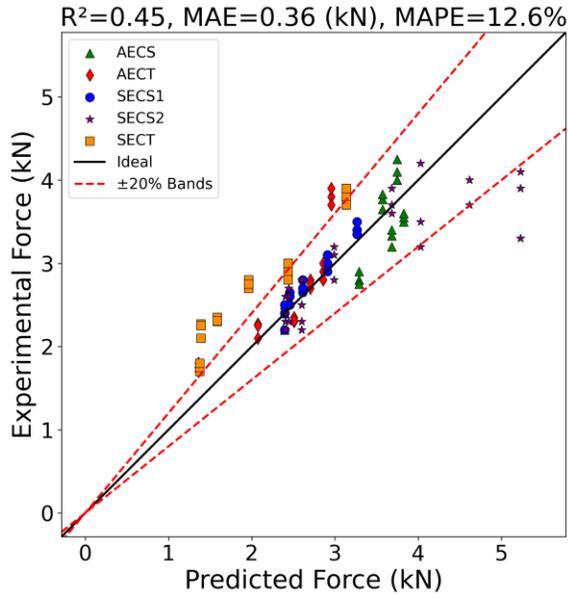
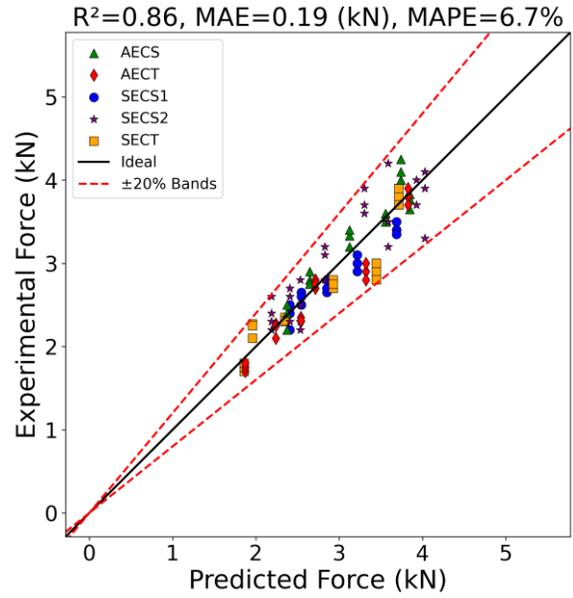

(a)                              (b)

**Fig. 6. Comparison of fracture load predictions: (a) GMTS model predictions and (b) MLP model predictions using stress-based inputs. The solid black line represents the ideal prediction, while the red dashed lines indicate ±20% scatter bands.**

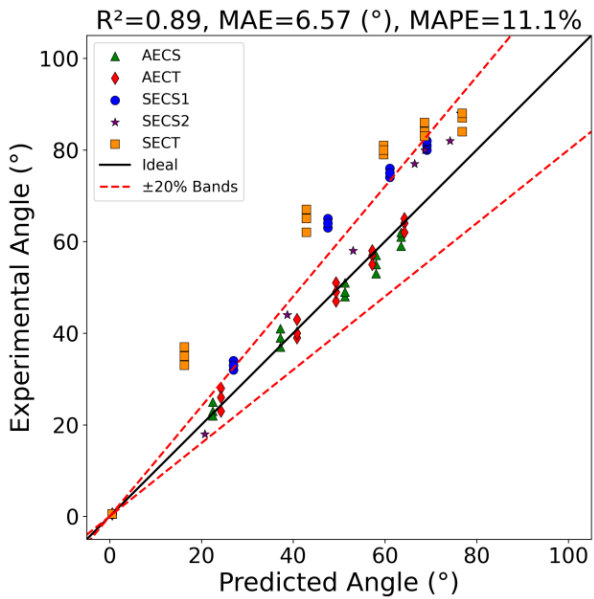
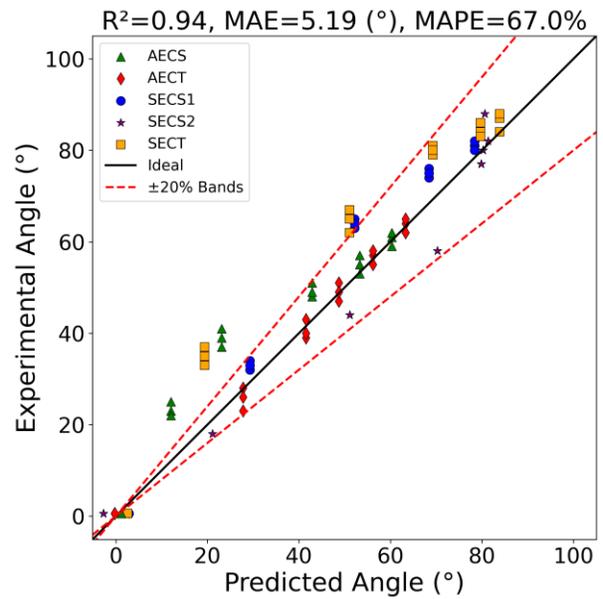

(a)                              (b)

**Fig. 7.** Comparison of fracture angle predictions: (a) GMTS model predictions and (b) MLP model predictions using stress-based inputs. The solid black line represents the ideal prediction, while the red dashed lines indicate ±20% scatter bands.

The GMTS model provided a reasonable baseline for predicting fracture behavior in certain scenarios, particularly for AECS and AECT specimens. However, as shown in Fig. 6(a), it struggled to consistently predict fracture load for SECT and SECS2 specimens, especially under mixed mode and Mode II (shear-dominated) conditions. This may be due to the fact that the critical distance is determined based on Mode I fracture toughness. For angle predictions, presented in Fig. 7, the GMTS model performed better overall compared to force predictions, with many data points falling within the scatter bands and aligning closely with the ideal line. Nevertheless, inconsistencies persisted across specimen types, particularly for SECT and SECS specimens, where several predictions deviated significantly and fell outside the scatter bands. However, in the case of SECT, the proposed framework also struggled to provide accurate predictions for all mode mixities. It is worth noting that the critical distance for SECS2 specimens was derived according to the original investigation and best-fitting procedures for all samples, rather than tested material properties. Even with this approach, the GMTS criterion exhibited lower accuracy compared to the proposed framework, particularly in addressing the complexities of mixed mode fracture behavior. To ensure a fair comparison, the outer-layer data removed by the algorithm were excluded from the calculations based on the GMTS criterion.

### 4.3. Discussion on feature selection algorithm

This section evaluates the effectiveness of the feature selection algorithm employed for stress-based predictions using an MLP model. By examining SHAP values, polar importance plots, and correlation heatmaps, presented in Fig. 8 to Fig. 10 respectively, this discussion provides insight into how these features interact and contribute to the model's predictive performance.

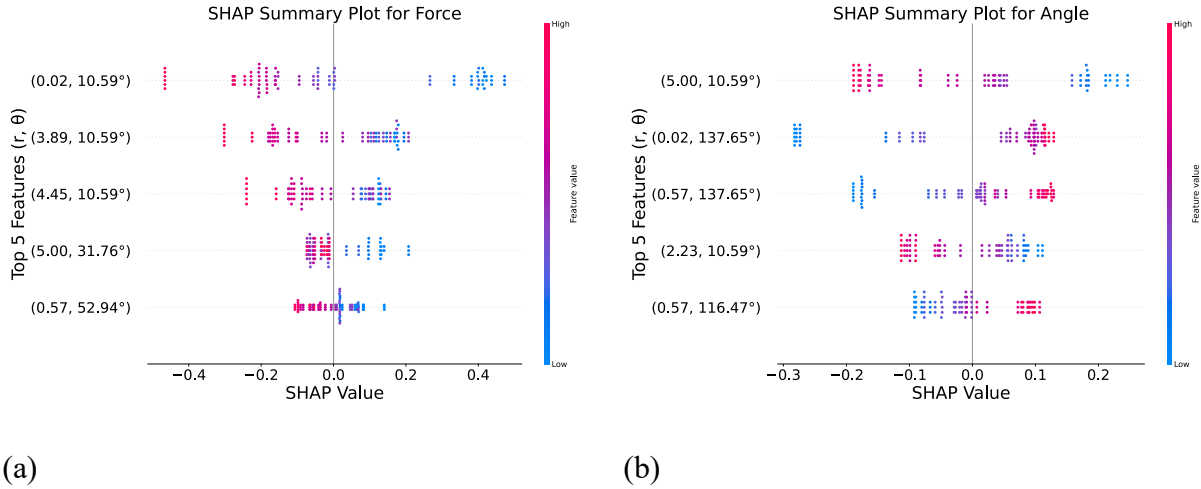

**Fig. 8. SHAP summary plot for the top 5 features presenting them in polar coordinates. (a) For fracture load predictions; (b) for fracture angle predictions.**

In Fig. 8, SHAP values are displayed on the x-axis and quantify the impact of each feature on the model's predictions. Each dot represents a single sample's feature value (colored by magnitude: red for higher stress, blue for lower) and its corresponding SHAP value. Positive SHAP values indicate an increase in the predicted quantity, while negative values indicate a decrease. The wide horizontal spread within each feature row underscores that feature's influence across the dataset. For fracture load, most influential features cluster at small angular positions near the crack tip, with some key points at larger radial distances (e.g., $r$=5.00 mm). These regions may correspond to stress fields extending beyond the immediate crack tip, reflecting the influence of the global stress distribution on overall fracture behavior, or they may reflect contributions from higher-order terms of the stress field and geometric constraints [55]. For fracture angle predictions, critical features span a broader angular range which illustrates the mixed mode (I/II) interactions. Overall, the algorithm successfully identifies physically meaningful features, bridging machine learning insights with fracture mechanics principles.

To simplify the interpretation of feature importance, the relative importance of features is visualized in a polar plot in Fig. 9, which provides a complementary perspective on the findings of the SHAP analysis.

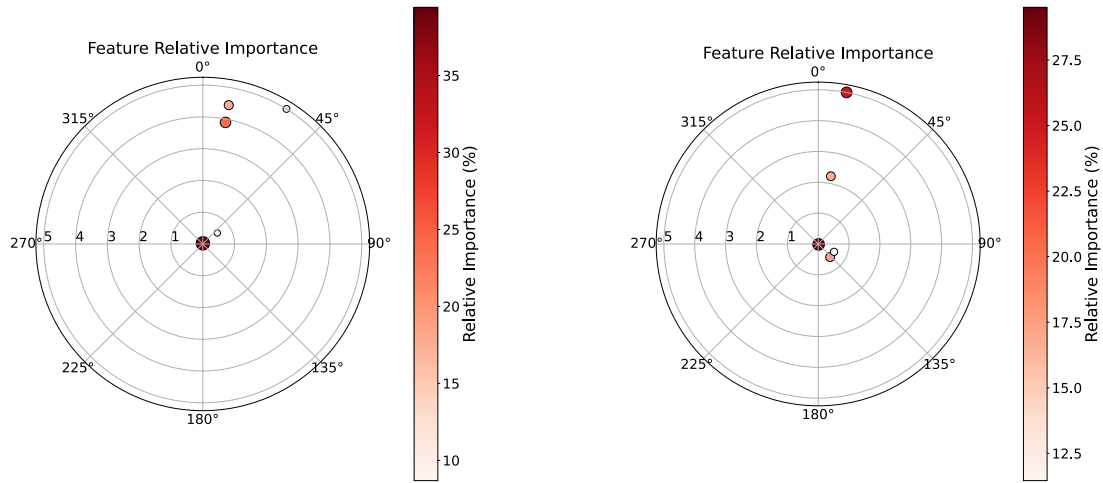

(a)                      (b)

**Fig. 9. Polar plot of feature relative importance for the top 5 features. (a) For fracture load predictions; (b) for fracture angle predictions.**

The polar plot highlights the dominance of features near the crack tip and at angular positions associated with high tensile stresses, which aligns with SHAP values as well as experimental results observed under mixed mode loading conditions. Finally, to expand on the analysis of the selected features, we present in Fig. 10 a heatmap showcasing the importance of all selected features for both fracture load and initiation angle predictions, providing a comprehensive view of their contributions to the machine learning model.

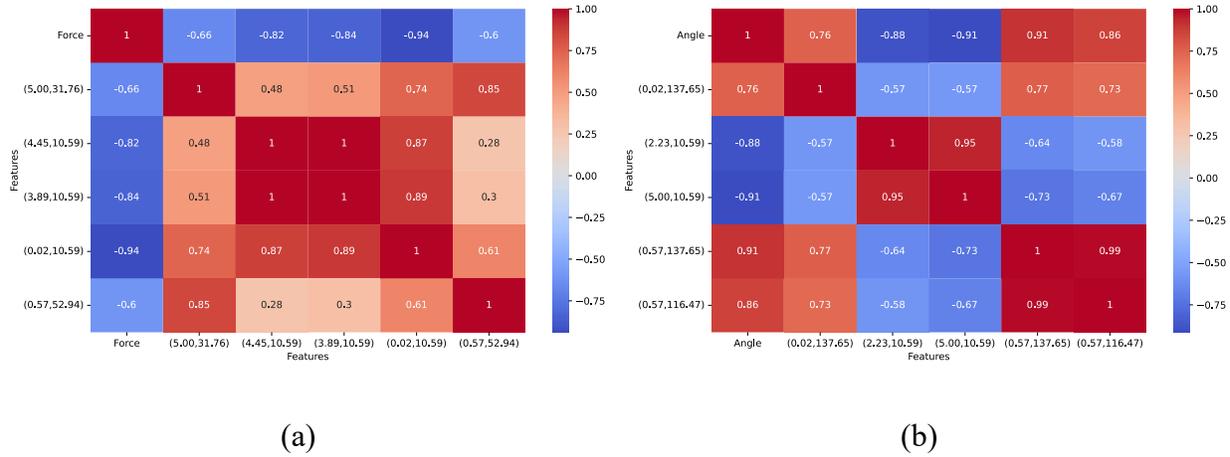

**Fig. 10. Feature correlation heatmap for the top 5 features presenting them in polar coordinates. (a) For fracture load predictions; (b) for fracture angle predictions.**

From Fig. 10 (as well as Fig. 9 and Fig. 8), it can be argued that the key radial distances, particularly 0.02 mm, exhibit the strongest intercorrelation with fracture load (correlation up to |0.94|) which highlights their physical significance in regions of high stress concentration near the crack tip. Larger radial distances, such as 5 mm, also correlate strongly with near-tip zones, while intermediate distances show near-unity mutual correlations, suggesting these features capture similar stress information. Meanwhile, ($r$=0.57 mm, $\theta$=52.94°) is moderately correlated with other features, implying a distinct mechanistic contribution beyond purely near-tip effects. For fracture angle, the heatmap highlights two main angular clusters around 10.59° and 116.47°–137.65° which reflect strong mixed mode or shear-tension interactions that guide the crack initiation angle. In summary, the presented SHAP plots, polar importance maps, and correlation heatmaps consistently indicate that near-tip stresses and certain far-field effects play pivotal roles in predicting fracture load and initiation angle. These findings reinforce the physical validity of the selected features and demonstrate how machine learning models integrate local and global stress information to yield accurate fracture predictions.

## 4.4. A discussion on selection of features

A critical consideration in implementing the framework lies in selecting representative nodes for the field distribution, such as the radial distance boundaries (start/end points), the number of radial sampling points, and angular divisions along the hoop axis. So far, we have used a radial range from 0.02 mm to 5 mm, with 10 radial points and 18 hoop divisions over 180°, totaling 180 points. To evaluate the framework's adaptability and performance, two distinct configurations are analyzed. Case 1 (Fig. 11): The radial distance spans from 0.04 mm to 2.5 mm, with 5 sampling points along the radial axis. The hoop axis is divided into 9 segments over 180°, resulting in a total of 45 data points (5 radial × 9 hoop). This configuration tests the framework's performance under sparse spatial sampling, evaluating its ability for fracture prediction with limited data. Case 2 (Fig. 12): The radial distance extends from 0.01 mm to 10 mm, discretized into 20 points along the radial axis. The hoop axis employs 36 divisions over 180°, yielding a total of 720 data points (20 radial × 36 hoop). This setup examines the framework's capacity to predict fracture load and angle using dense spatial data. This case specifically demonstrates the efficiency of the feature selection algorithm in identifying critical features from a large dataset. Note that the baseline configuration (radial: 0.02–5 mm, 10 points; hoop: 18 divisions) can be used as a reference for intermediate complexity. By comparing these setups, the analysis highlights how parameter selection influences the prediction accuracy of the proposed approach.

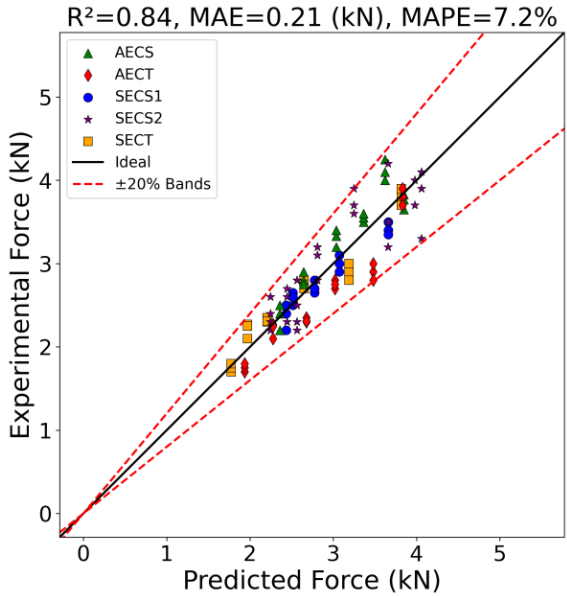 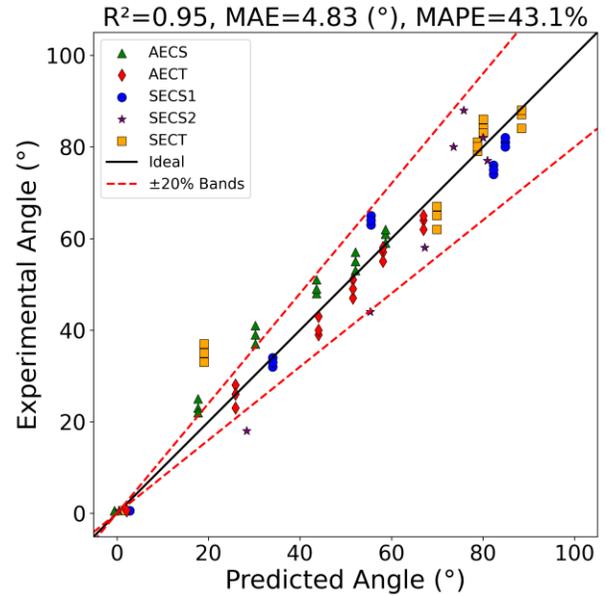

(a)                        (b)

**Fig. 11. Predictions of fracture load (a) and initiation angle (b) for a configuration with a radial range from 0.04 mm to 2.5 mm, 5 points along the radial axis, and 9 points along the hoop axis (total: 45 features). The solid black line represents the ideal prediction, while the red dashed lines indicate ±20% scatter bands.**

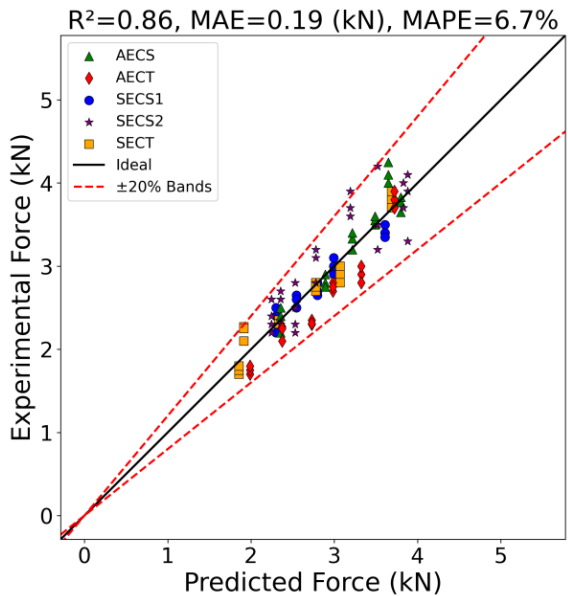 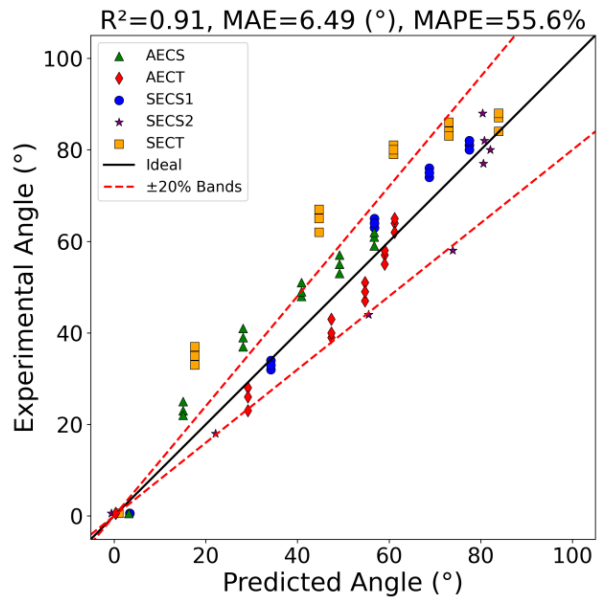

(a)                        (b)

**Fig. 12.** Predictions of fracture load (a) and initiation angle (b) for a configuration with a radial range from 0.01 mm to 10 mm, 20 points along the radial axis, and 36 points along the hoop axis (total: 720 features). The solid black line represents the ideal prediction, while the red dashed lines indicate ±20% scatter bands.

In Fig. 11, with a reduced number of features, as well as Fig. 12 with a higher number of features, the framework provides almost the same accuracy for both fracture load and initiation angle predictions. These results highlight the robustness of the framework and its ability to accurately predict fracture across a wide range of feature densities, from 45 to 720 features. They also demonstrate the effectiveness of the feature selection algorithm in identifying the most relevant features. The proposed framework employs a thorough, three-step feature selection process that significantly enhances its predictive capability. First, Mutual information analysis quantifies the dependency between each candidate feature and the fracture outcomes, ensuring that only those with substantial informational value are used. Next, the Lasso-based feature reduction step applies *L1* regularization to enforce sparsity, which eliminates redundant or less influential features and enhances the model's efficiency. Finally, SHAP analysis interprets the contribution of each feature by assigning importance scores based on Shapley values, which confirms the relevance and impact of the selected features on fracture load and initiation angle predictions.

**4.5. Limited data for training**

To further demonstrate the efficiency and robustness of the proposed framework and to show that it extends beyond a simple curve-fitting tool with limited generalizability, the model is trained using only pure Mode I and Mode II data (limited training conditions), which represent approximately one-third of the entire dataset (a limited dataset, roughly 30 samples). The model's predictions are then evaluated on the remaining data, all of which correspond to mixed mode I/II conditions as unseen data. The results are presented in Fig. 13.

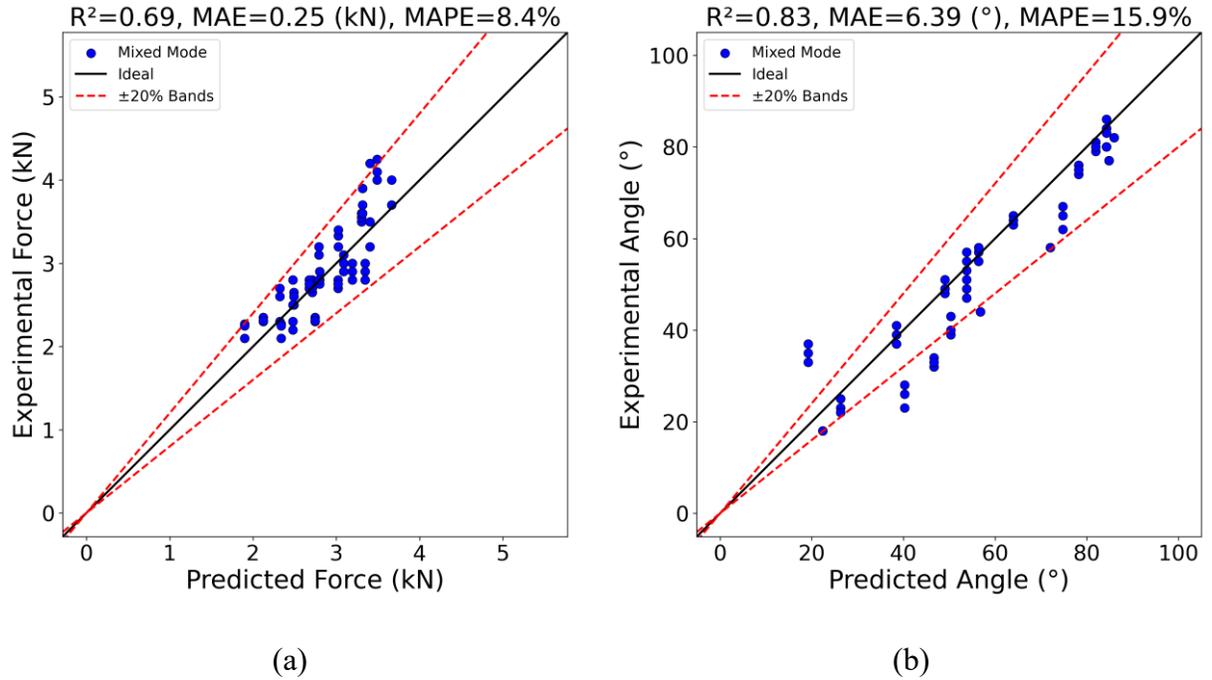

(a)                                         (b)

**Fig. 13.** Evaluation of the proposed framework's predictions for fracture load and initiation angle after training exclusively on pure Mode I and Mode II data. The solid black line represents the ideal prediction, while the red dashed lines indicate ±20% scatter bands.

The results in Fig. 13 demonstrated that the proposed framework delivers highly accurate predictions, with almost all fracture load predictions falling within the ±20% scatter band and achieving good accuracy for fracture initiation angle, despite the lack of training on mixed mode loading conditions. The framework performed comparably to the GMTS model while achieving higher $R^2$ and lower MAE values in both fracture load and angle predictions. It is worth recalling that, in the case of SECS2, except for pure mode I and II loading conditions, there are six distinct mixed mode configurations. These results highlight the framework's efficiency and its capability to predict fracture behavior under complex loading conditions. Furthermore, they emphasize the framework's ability to generalize effectively, capturing key physical trends and offering a robust alternative to traditional analytical models for mixed mode fracture predictions.

Overall, the presented ML framework represents a significant advancement in the predictive modeling of fracture behavior. It has been shown to be efficient and accurate. Future work could explore applying this framework to more complex material behavior and loading conditions, as well as integrating it into real-time monitoring and predictive maintenance systems for engineering structures. Despite its promising results, the proposed framework has some limitations that warrant consideration. From a numerical perspective, the framework's accuracy depends on the quality and variety of the training data, which is a typical characteristic of machine learning methods. Furthermore, while the feature selection process using Mutual Information, Lasso, and SHAP is robust, it may overlook subtle interactions between features, particularly when dealing with highly correlated datasets or noisy input data. The dataset used in this study includes only one material; thus, assessing the performance of the proposed framework on other materials would be valuable. Nevertheless, the dataset used in this study comprises 100 unique experimental data points, covering 32 different cases across five distinct cracked specimens under various mode mixities, offering both fracture load and angle information. It should be noted that, due to the inherent measurement errors and data scatter present in the experimental data, the results support the model's ability to describe behavior under noisy conditions.

## 5. Conclusions

This study introduced a machine learning framework for predicting mixed mode I/II fracture toughness and crack initiation angles in brittle materials. By using stress, strain, or displacement field distributions represented by nodal data around the crack tip, the framework achieved high predictive accuracy across various mode mixities and specimen geometries. A key distinguishing feature of this work is its reliance on the unique field distributions around the crack tip, which is specific to each mode mixity and specimen geometry. This approach offers a direct and effective means of assessing failure without the need for post-processing steps, such as stress intensity factor calculations. As a result, the framework is both easy to implement and computationally efficient.

By evaluating the model's performance across a wide range of experimental data, including five different specimens and at least six mode mixities per specimen, this study demonstrated that among stress, strain, and displacement inputs, stress-based features consistently yielded the highest accuracy for both fracture load and initiation angle predictions across various ML models, while the other inputs also demonstrated strong predictive performance. The predictions achieved an $R^2$ of 0.86 for fracture load and 0.94 for crack initiation angle when using a multilayer perceptron (MLP) model with stress-based inputs. These results demonstrated superior performance even compared to the most well-established and accurate theoretical models, such as TCD (GMTS).

To handle the high-dimensional input space or scenarios involving sparse or low-dimensional data, a feature selection pipeline combining Mutual Information analysis, Lasso regression, and SHAP-based interpretability techniques was implemented. The selected features were shown to be statistically significant and physically meaningful, with nodes near the crack tip consistently identified as the most important. Moreover, the framework highlighted the importance of far-field components, which reflect the influence of the global stress distribution on fracture behavior. The study also evaluated the framework's adaptability to limited datasets as well as restricted mode mixity scenarios, which resulted in accurate predictions and confirmed its practical applicability in such cases. Because of the model's high predictive capability when using strain or displacement fields, it can potentially be used for in-situ failure assessments by integrating digital image correlation (DIC) techniques.

## References

[1] T.L. Anderson, Fracture mechanics: fundamentals and applications, CRC press, 2005.

[2] A.A. Griffith, VI. The phenomena of rupture and flow in solids, Philos. Trans. R. Soc. London. Ser. A, Contain. Pap. a Math. or Phys. Character 221 (1921) 163–198.

[3] F. Erdogan, G.C. Sih, On the Crack Extension in Plates Under Plane Loading and Transverse Shear, J. Basic Eng. 85 (1963) 519–525editor & translator.

[4] G.C. Sih, Strain-energy-density factor applied to mixed mode crack problems, Int. J. Fract. 10 (1974) 305–321editor & translator.

[5] K.J. Chang, On the maximum strain criterion—a new approach to the angled crack problem, Eng. Fract. Mech. 14 (1981) 107–124.

[6] D. Gope, P.C. Gope, A. Thakur, A. Yadav, Application of artificial neural network for predicting crack growth direction in multiple cracks geometry, Appl. Soft Comput. 30 (2015) 514–528editor & translator.

[7] M.R. Ayatollahi, M. Rashidi Moghaddam, F. Berto, T-stress effects on fatigue crack growth – Theory and experiment, Eng. Fract. Mech. 187 (2018) 103–114editor & translator.

[8] R. Ince, Prediction of fracture parameters of concrete by Artificial Neural Networks, Eng. Fract. Mech. 71 (2004) 2143–2159editor & translator.

[9] X. Li, X. Zhang, W. Feng, Q. Wang, Machine learning-based prediction of fracture toughness and path in the presence of micro-defects, Eng. Fract. Mech. 276 (2022) 108900editor & translator.

[10] D.A. Hills, P.A. Kelly, D.N. Dai, A.M. Korsunsky, Solution of crack problems: the distributed dislocation technique, Springer Science \& Business Media, 2013.

[11] R. Bagheri, A.M. Mirzaei, Fracture Analysis in an Imperfect FGM Orthotropic Strip Bonded Between Two Magneto-Electro-Elastic Layers, Iran. J. Sci. Technol. - Trans. Mech. Eng. 43 (2019)editor & translator.

[12] A. Seibi, S.M. Al-Alawi, Prediction of fracture toughness using artificial neural networks (ANNs), Eng. Fract. Mech. 56 (1997) 311–319editor & translator.

[13] K. Zarrabi, K.H. Tsang, An elastic mixed-modes I and II fracture criterion using an artificial neural network database, (2007).

[14] B. Bahrami, H. Talebi, M.R. Ayatollahi, M.R. Khosravani, Artificial neural network in prediction of mixed-mode I/II fracture load, Int. J. Mech. Sci. 248 (2023) 108214editor & translator.

[15] X. Liu, C.E. Athanasiou, N.P. Padture, B.W. Sheldon, H. Gao, A machine learning approach to fracture mechanics problems, Acta Mater. 190 (2020) 105–112editor & translator.

[16] E. Emami Meybodi, S.K. Hussain, M. Fatehi Marji, V. Rasouli, Application of machine

learning models for predicting rock fracture toughness mode-I and mode-II, J. Min. Environ. 13 (2022) 465–480.

[17] A. Mahmoodzadeh, D. Fakhri, A. Hussein Mohammed, A. Salih Mohammed, H. Hashim Ibrahim, S. Rashidi, Estimating the effective fracture toughness of a variety of materials using several machine learning models, Eng. Fract. Mech. 286 (2023) 109321editor & translator.

[18] A. Bagher Shemirani, Prediction of fracture toughness of concrete using the machine learning approach, Theor. Appl. Fract. Mech. 134 (2024) 104749editor & translator.

[19] A. Eskandariyun, A. Joseph, A. Stere, A. Byar, S. Fomin, M. Kabir, J. Dong, N. Zobeiry, A Combined Finite Element and Machine Learning Approach to Accelerate Calibration and Validation of Numerical Models for Prediction of Failure in Aerospace Composite Parts, (2023)editor & translator.

[20] S. Goswami, C. Anitescu, S. Chakraborty, T. Rabczuk, Transfer learning enhanced physics informed neural network for phase-field modeling of fracture, Theor. Appl. Fract. Mech. 106 (2020) 102447editor & translator.

[21] B. Zhu, H. Li, Q. Zhang, Extended physics-informed extreme learning machine for linear elastic fracture mechanics, Comput. Methods Appl. Mech. Eng. 435 (2025) 117655editor & translator.

[22] P. Carrara, L. De Lorenzis, L. Stainier, M. Ortiz, Data-driven fracture mechanics, Comput. Methods Appl. Mech. Eng. 372 (2020) 113390editor & translator.

[23] H. Talebi, B. Bahrami, H. Ahmadian, M. Nejati, M.R. Ayatollahi, An investigation of machine learning algorithms for estimating fracture toughness of asphalt mixtures, Constr. Build. Mater. 435 (2024) 136783editor & translator.

[24] A.M. Mirzaei, Stress, Strain, or Energy? which one is superior predictor of fatigue life in notched Components? a novel Machine Learning-Based framework, Eng. Fract. Mech. 309 (2024) 110401editor & translator.

[25] I. Yeo, R.A. Johnson, A new family of power transformations to improve normality or symmetry, Biometrika 87 (2000) 954–959editor & translator.

[26] M.L. Williams, Stress singularities resulting from various boundary conditions in angular corners of plates in extension, J. Appl. Mech. ASME 19 (1952) 526–528.

[27] G.A.F. Seber, A.J. Lee, Linear regression analysis, John Wiley \& Sons, 2012.

[28] P.J. Huber, Robust estimation of a location parameter, in: Break. Stat. Methodol. Distrib., Springer, 1992: pp. 492–518.

[29] C.E. Shannon, A mathematical theory of communication, Bell Syst. Tech. J. 27 (1948) 379–423.

[30] R. Tibshirani, Regression Shrinkage and Selection Via the Lasso, J. R. Stat. Soc. Ser. B 58 (2018) 267–288editor & translator.

[31] T. Chen, C. Guestrin, XGBoost: A Scalable Tree Boosting System, in: Proc. 22nd ACM

SIGKDD Int. Conf. Knowl. Discov. Data Min., Association for Computing Machinery, New York, NY, USA, 2016: pp. 785–794editor & translator.

[32] J. Von Neumann, O. Morgenstern, Theory of games and economic behavior: 60th anniversary commemorative edition, in: Theory Games Econ. Behav., Princeton university press, 2007.

[33] M. Stone, Cross-validatory choice and assessment of statistical predictions, J. R. Stat. Soc. Ser. B 36 (1974) 111–133.

[34] J. Bergstra, R. Bardenet, Y. Bengio, B. Kégl, Algorithms for hyper-parameter optimization, Adv. Neural Inf. Process. Syst. 24 (2011).

[35] F. Rosenblatt, The perceptron: A probabilistic model for information storage and organization in the brain., Psychol. Rev. 65 (1958) 386–408editor & translator.

[36] D.E. Rumelhart, G.E. Hinton, R.J. Williams, Learning representations by back-propagating errors, Nature 323 (1986) 533–536editor & translator.

[37] N. Srivastava, G. Hinton, A. Krizhevsky, I. Sutskever, R. Salakhutdinov, Dropout: a simple way to prevent neural networks from overfitting, J. Mach. Learn. Res. 15 (2014) 1929–1958.

[38] T.K. Ho, Random decision forests, in: Proc. 3rd Int. Conf. Doc. Anal. Recognit., 1995: pp. 278–282 vol.1editor & translator.

[39] L. Breiman, Random Forests, Mach. Learn. 45 (2001) 5–32editor & translator.

[40] L. Breiman, Bagging predictors, Mach. Learn. 24 (1996) 123–140.

[41] A. Natekin, A. Knoll, Gradient boosting machines, a tutorial, Front. Neurorobot. 7 (2013)editor & translator.

[42] J.H. Friedman, Greedy Function Approximation: A Gradient Boosting Machine, Ann. Stat. 29 (2001) 1189–1232. http://www.jstor.org/stable/2699986 (accessed July 9, 2023).

[43] L.G. Valiant, A theory of the learnable, Commun. ACM 27 (1984) 1134–1142.

[44] S. Lundberg, A unified approach to interpreting model predictions, ArXiv Prepr. ArXiv1705.07874 (2017).

[45] F. Pedregosa, G. Varoquaux, A. Gramfort, V. Michel, B. Thirion, O. Grisel, M. Blondel, P. Prettenhofer, R. Weiss, V. Dubourg, others, Scikit-learn: Machine learning in Python, J. Mach. Learn. Res. 12 (2011) 2825–2830.

[46] A.M. Mirzaei, M.R. Ayatollahi, B. Bahrami, F. Berto, A new unified asymptotic stress field solution for blunt and sharp notches subjected to mixed mode loading, Int. J. Mech. Sci. 193 (2021) 106176editor & translator.

[47] D.J. Smith, M.R. Ayatollahi, M.J. Pavier, The role of T-stress in brittle fracture for linear elastic materials under mixed-mode loading, Fatigue Fract. Eng. Mater. Struct. 24 (n.d.) 137–150editor & translator.

[48] M.R.M. Aliha, A. Bahmani, S. Akhondi, Mixed mode fracture toughness testing of

PMMA with different three-point bend type specimens, Eur. J. Mech. - A/Solids 58 (2016) 148–162editor & translator.

[49] M.R. Ayatollahi, M.R.M. Aliha, M.M. Hassani, Mixed mode brittle fracture in PMMA—An experimental study using SCB specimens, Mater. Sci. Eng. A 417 (2006) 348–356editor & translator.

[50] H. Neuber, Theory of notch stresses: principles for exact calculation of strength with reference to structural form and material., (1961) 293p. file://catalog.hathitrust.org/Record/102014224.

[51] D. Taylor, The theory of critical distances, Eng. Fract. Mech. 75 (2008) 1696–1705editor & translator.

[52] D. Taylor, M. Merlo, R. Pegley, M.P. Cavatorta, The effect of stress concentrations on the fracture strength of polymethylmethacrylate, Mater. Sci. Eng. A 382 (2004) 288–294editor & translator.

[53] A.M. Mirzaei, M.R. Ayatollahi, B. Bahrami, F. Berto, Elastic stress analysis of blunt V-notches under mixed mode loading by considering higher order terms, Appl. Math. Model. 78 (2020) 665–684editor & translator.

[54] R.K. Nalla, J.H. Kinney, R.O. Ritchie, Mechanistic fracture criteria for the failure of human cortical bone, Nat. Mater. 2 (2003) 164–168.

[55] B. Bahrami, M.R. Ayatollahi, A.M. Mirzaei, F. Berto, Improved stress and displacement fields around V-notches with end holes, Eng. Fract. Mech. 217 (2019) 106539editor & translator.